\documentclass{vgtc}                          

\ifpdf
  \pdfoutput=1\relax                   
  \usepackage{graphicx}                
  \DeclareGraphicsExtensions{.pdf,.png,.jpg,.jpeg} 
\else
  \usepackage{graphicx}                
  \DeclareGraphicsExtensions{.eps}     
\fi%

\graphicspath{{figures/}{pictures/}{images/}{./}} 

\usepackage{microtype}                 
\PassOptionsToPackage{warn}{textcomp}  
\usepackage{textcomp}                  
\usepackage{mathptmx}                  
\usepackage{times}                     
\usepackage{cite}                      
\usepackage{tabu}                      
\usepackage{booktabs}                  

\usepackage{hyperref}
\hypersetup{
    colorlinks=true,
    linkcolor=blue,
    filecolor=magenta,      
    urlcolor=cyan,
}
\usepackage{csquotes}

\onlineid{0}

\vgtccategory{Research}

\vgtcinsertpkg

\title{A Psychology of Visualization or (External) Representation?}

\author{Amy Rae Fox\thanks{e-mail: amyraefox@ucsd.edu} 
}
\affiliation{\scriptsize Department of Cognitive Science \\ University of California, San Diego}
        
\abstract{ 
What \textit{is} a visualization? There is limited utility in trifling with definitions, except insofar as one serves as a tool for communicating and conceptualizing our subject matter;  a statement of identity for a community. To establish Visualization Psychology as a viable interdisciplinary research programme, we must first define the object(s) of our collective inquiry. I propose that while we might refer to the study of ``visualization" for the term’s colloquial accessibility and pragmatic alignment with other fields, we should consider for exploration a class of artifacts and corresponding processes more expansive and profound: external representations. What follows is an argument for the study of external representation as the foundation for a new interdisciplinary endeavor, and approach to mapping the corresponding problem space. 
}

\CCScatlist{ 
 information visualization, external representation
}

\begin{document}



\maketitle

\section{Introduction}

The language of representation is slippery and self-referencing. If I show you a collection of artifacts containing marks on surfaces, you might label some as pictures or art, others as diagrams, maps, schematics, some graphs, charts or plots, and others also graphs but you might use air quotes and call them ``graph-theory graphs". Some you'll identify as writing, and others, like writing but not writing\textemdash some peculiar or particular systems of notation. The labels you apply to each marking likely depend on your disciplinary background, and are neither exhaustive, nor mutually exclusive. Which of these, are visualizations?

\section{On Visualization}

Let us start with definitions put forth in popular Visualization texts. Stephen Few offers a functional definition, characterizing data
visualization as \textit{``an umbrella term to cover all types of visual representations that support the exploration, examination, and communication of data. Whatever the representation, as long as it’s visual, and whatever it represents, as long as it’s information, this constitutes data visualization"} \cite[pg.12]{Few2009}. This is a delightfully inclusive specification, according to which the words on this page would constitute a visualization\textemdash but would scarcely be considered so by most visualization practitioners. Why? Because visualizations are somehow \textit{graphic} in nature, more depictive than descriptive. From Ware \cite[pg.2]{Ware2004},\textit{``a graphical representation of data or concepts."} In fact, one might find it easier to depict the set of markings one considers visualizations than to describe them in words. Similarly, in their timeless text, Card, Mackinlay and Shneiderman define information visualization as, \textit{``The use of computer-supported, interactive, visual representations of abstract data to amplify cognition"} \cite[pg.7]{Card1999}. In both cases, we see the (appropriate) characterization of information visualization as artifact \textit{and} process. But we are left with an under-specification of what constitutes such an artifact. 
Need the representations be interactive, computer-generated marks on \textit{surfaces}? Is the nature of the information constrained?
What makes a marking graphic? I draw on these examples not in critique of their notable contributions, but rather to call attention to a gap in the foundation of the field, and subsequent opportunity for psychologists (and perhaps philosophers) of visualization to make an impactful contribution. 

Definitions, as terminology, serve as tools for communicating and conceptualizing one's subject matter\cite{Cabre1999}. I argue that to establish Visualization Psychology as a viable interdisciplinary research programme, we must first define the object(s) of our collective inquiry. To this I propose that while we might prefer the term \textit{visualization} for its colloquial accessibility and pragmatic alignment with the SCI/VIS/VAST communities, there is an alternative characterization of artifacts and processes more pertinent to the way ``visualizations" are encountered in everyday life: as external representations. 

\section{On External Representation}
The term \textit{external representation} stems from early cognitive science and information-processing psychology; inquiries into the existence and nature of mental representations (see \cite{Neisser,LindsayNorman72,PalmerKimchi86}). Palmer argued that as cognitive representations are, ``exceedingly complex and difficult to study," one might start with the examination of ``noncognitive" (ie. external) representations, as they are ``simple, and easy to study"\footnote{More "accessible" being perhaps the more accurate characterization.} \cite[pg.262]{Palmer1978}. His subsequent elaboration of \textit{representational systems} demonstrates there is much to explore with respect to the nature and function of such ``noncognitive" structures, without reliance on the form of any internal counterparts. Zhang \& Norman described external representations as knowledge and structure, \textit{``in the world, as physical symbols (e.g., written symbols, beads of abacuses, etc.) or as external rules, constraints, or relations embedded in physical configurations (e.g., spatial relations of written digits, visual and spatial layouts of diagrams, physical constraints in abacuses, etc.)}" \cite[pg.3]{ZhangNorman1994}. Zhang later writes that the information \textit{in} external representations, \textit{``can be picked up, analyzed, and processed by perceptual systems alone, although the top-down participation of conceptual knowledge from internal representations can sometimes facilitate or inhibit the perceptual processes."}\cite[pg.180]{Zhang1997}.  What is wisely made explicit in these characterizations is the assertion that external representations do not exist in isolation, rather, they work in concert with internal representations, whatever their form. 

In the proceeding decades, cognitive scientists took up the challenge of discovering how various forms of external representation influence various forms of thinking; with particular attention to the fashion in which representations support computation, such as in problem solving \cite{Larkin1987}, and scientific discovery \cite{ChengSimon95}. Distinctions were drawn between the sentential/propositional, and graphic/diagrammatic, where the latter class was taken up by its own interdisciplinary community in the early 2000s.
Educational psychologists and learning scientists turned their attention to multimodal representations (where modality refers both to the sensory modality (eg. visual, auditory) as well as the encoding media \cite{SchnotzIMTPC, MayerCTML}. By the late 2000s, sufficient interest across allied disciplines warranted a special issue of the journal TopiCS in Cognitive Science dedicated to visual-spatial representations, with milestone contributions on visual analytics \cite{Fisher2011}, graph comprehension \cite{ShahFreedman2011}, and diagrams \cite{Cheng2011}, with comprehensive reviews of visual-spatial representations as tools for thinking\cite{Tversky2011} and corresponding implications for design \cite{Hegarty2011}. It is worth noting that while the articles discussing visual analytics, diagrams and statistical graphics might have found a home at a VIS conference (after the 2010 cognitive turn) the latter review articles would be more out of place, but are indicative of work that should be at the theoretical core of a Visualization Psychology. 

\section{A Pragmatic Proposal}

As psychologists, we are concerned not only with the tools, design, and efficacy of such representations, but with their mechanics: how they function (or not). This function is enacted between the artifact(s) and person(s), embodied, and situated in their environments and complex social structures. What follows is a first proposal for what minimal conception of external representation might be taken into the ``hard core" of any Lakatosian research programme in Visualization Psychology. 

\vspace{-1mm}
\begin{quote}
    When we study a visualization, we are studying the function of an external representation: the construction of meaning in a distributed cognitive system. The construction of meaning is oft followed by intelligent action with that meaning, be it by learning, making a decision, solving a problem, forming a judgement, or any of a multitude of complex cognitive activities which form the \textit{communicative context} of the representation.
\end{quote}
\vspace{-1mm}

Here we admit visualization as a subset of external representation, an active construction (rather than transmission) of meaning, and that meaning serves some purpose in the context of interaction with the representation. What is crucial is that we orient ourselves equally toward the artifact \textit{and} the procedure; representation as \textit{thing}, and representation as \textit{process}. 

\section{A Problem Space}

We have moved from the study of computer-generated, interactive, graphics, to any externalization of thought. What we are left with, it seems, is a Goldilocks problem. The idiomatic conception of visualization is too narrow, and a faithful conception of external representation, too broad. Fortunately there are dimensions along which this metaphorical problem space can be surveyed. We might think of these dimensions as ranges along which we can tune our attention, progressively expanding or narrowing our scope of inquiry depending on the state of theoretical and technological advancement. 

\subsection{On Sensory Modality: How we represent}

External representations can be constructed for any sensory modality, though by far the most attention has been paid to the visual. Deservedly so, as visuals are the most pervasive information artifacts, and the sensory modality about which we have the most understanding. Though we are surely far from exhausting the wellspring of questions to be asked of visual representations, and contributions for vision scientists to make, I suggest that we accept within our scope multi-sensory representations. From a theoretical stance, this requires broader inclusion of expertise across perceptual psychology, though the applications are consequential. Accessibility demands informationally-equivalent representations for those without visual perception, in an increasingly visualization-driven world. 

\subsection{On Encoding: How we represent}
Though I've noted the lack of precision in defining the scope of visualizations, there has been no lack of effort in cataloging \cite{Harris1990} and taxonomizing them, from general descriptive frameworks\cite{Tan1990,Shneiderman1996,Chi2000,Tory2004,ParsonsSedig2014a,Blackwell2002}, to those concerned with specific domains of data \cite{Aigner2007,Beck2016,Blascheck2017}. Two particularly useful (and under-appreciated) are those of Engelhardt \cite{Engelhardt2002} who offers an atomic, generative framework deserving of its characterization as a \textit{language} of graphics, and Massironi \cite{Massironi} who offers both a taxonomy and evolutionary timeline. 

While most taxonomies deal with some intersection of graphical structure and data type (eg. geographic-maps, relational-networks), the more common distinction in the cognitive and learning science literature is the continuum from \textit{descriptive} to \textit{depictive}, roughly analogous with symbolic to analog, or propositional to graphic. These terms refer to a semiotic modality (also: medium), which indicates the degree of convention (how arbitrary) the relation between a representation and thing to which it refers.  While the poles of a depictive--descriptive continuum can be easily identified, there lays betwixt a murky medium. At what point of abstraction does an icon become a symbol? When it is no longer identifiable as its referent without convention? In whose judgment? We are more accurate in describing our scope of inquiry as multimedia, than `primarily graphic'. I propose that while origins of visualization as a field lie in the distinction of graphics from text,  
fundamental questions about framing, persuasion, and even comprehension rely on understanding the function of text alongside graphics. It is rarely the case that external representations of the visual graphic variety are not accompanied by some form of linguistic propositions or notation. Indeed, a visualization without a title and labels may be worth no words at all. 

\subsection{On Purpose: Why we represent}
Visualization texts describe the purpose of visualization as being to `amplify' cognition \cite{Few2009, Ware2004, Card1999}, though as psychologists we appreciate the story is more nuanced \cite{Kirsh2010}. What kind of cognition, to what end? The most generic case is that of communicating to simply \textit{inform}: the boxplot in my manuscript or barchart in newspaper, where I aim  to inform the reader of some aspects of the underlying information, in as clear a manner as possible. But I might design that artifact differently if I want you to explore the data, to undertake an analysis, make a decision, a plan, or a forecast. I'll certainly change my strategy if I want to strongly persuade you, or alternatively, want you to use the representation to learn. There are entire systems of diagrams designed for solving particular kinds of problems, and the design of representations to support conceptual change is the focus of entire subdiscplines in STEM education.  I use the term \textit{communicative context}, to refer to the ``cognitive activity" the designer of a representation intends the user to perform. The structure of these activities has not been taxonomized, though a compelling framework for the hierarchical, emergent structure of such activities is detailed by Sedig \& Parsons \cite{SedigParsonsVA2012, SedigParsonsMacro}. The relevant insight  is that certain parameters of a representation, such as the computational efficiency, or relative explicitness of certain aspects of the data, will be tuned in accordance with the task the reader is expected to perform.

\section{Conclusion}

We began with the question, `What is a visualization?' and end with the contention it is a class too narrow to characterize the scope of our endeavor. When we study visualizations in real-world activity, we are actually studying multimedia representations. Similarly, we must work at the development and understanding of multimodal multimedia representations in order to realize a world where access to information is not the exclusive privilege of those with visual perception. While we might focus primarily on graphic visuals to the extent that they are the most common tools for representing information, these porous boundaries are both pragmatic and realistic. By adopting a broader scope, our nascent programme positions itself at the boundaries of established visualization scholarship, alongside researchers exploring data visceralization, accessibility and multimodal communication, multimedia learning and a wider community of scholars who share interest in diverse communicative artifacts.


\bibliographystyle{abbrv-doi}

\bibliography{template}
\end{document}